\documentclass[electronics,article,accept,pdftex,moreauthors]{Definitions/mdpi} 
\firstpage{1} 
\pubvolume{15}
\issuenum{15}
\articlenumber{769}
\pubyear{2026}
\copyrightyear{2026}
\externaleditor{Xin Luo} % More than 1 editor, please add `` and '' before the last editor name
\datereceived{26 December 2025} 
\daterevised{26 January 2026} % Comment out if no revised date
\dateaccepted{9 February 2026} 
\datepublished{11 February 2026} 

\Title{A Data-Driven Multimodal Method for Early Detection of Coordinated Abnormal Behaviors in Live-Streaming Platforms}

\Author{{Jingwen Luo} %MDPI: Please carefully check the accuracy of names and affiliations.
 $^{1,2,\dagger}$, Pinrui Zhu $^{1,\dagger}$, Yiyan Wang $^{1,\dagger}$, Zilin Xiao $^{1,3,\dagger}$, Jingqi Li $^{1}$, Xuebei Kong $^{1}$ and Yan Zhan $^{1,4,}$*}

\AuthorNames{Jingwen Luo, Pinrui Zhu, Yiyan Wang, Zilin Xiao, Jingqi Li, Xuebei Kong, Yan Zhan}

\address{%
$^{1}$ \quad {National School of Development, Peking University}%MDPI: The department/school/faculty/campus of this university is required. Please try to provide this information and check all affs. 
, Beijing 100871, China %MDPI: We added these email addresses here according to those submitted online at susy.mdpi.com. Please confirm.
\\
$^{2}$ \quad {College of Information and Electrical Engineering, China Agricultural University}, Beijing 100083, China\\
$^{3}$ \quad {School of Foreign Studies, Beijing Language and Culture University}, Beijing 100083, China\\
$^{4}$ \quad Artificial Intelligence Research Institute, Tsinghua University, Beijing 100084, China \\
}

\corres{Correspondence: yan\_zhan\_thu@163.com}
\firstnote{These authors contributed equally to this work.} 

\usepackage{changes}
\usepackage{verbatim}
\usepackage{algorithm}
\usepackage{algorithmic}
\usepackage{amsfonts, amssymb}
\usepackage{gensymb}
\usepackage{bbding}
\usepackage{makecell}

\abstract{\textls[-15]{With the rapid growth of live-streaming e-commerce and digital marketing, abnormal marketing behaviors have become increasingly concealed, coordinated, and intertwined across heterogeneous data modalities, posing substantial challenges to data-driven platform governance and early risk identification. Existing approaches often fail to jointly model cross-modal temporal semantics, the gradual evolution of weak abnormal signals, and organized group-level manipulation. To address these challenges, a data-driven multimodal abnormal behavior detection framework, termed MM-FGDNet, is proposed for large-scale live-streaming environments. The framework models abnormal behaviors from two complementary perspectives, namely temporal evolution and cooperative group structure. A cross-modal temporal alignment module first maps video, text, audio, and user behavioral signals into a unified temporal semantic space, alleviating temporal misalignment and semantic inconsistency across modalities. Building upon this representation, a temporal fraud pattern modeling module captures the progressive transition of abnormal behaviors from early incipient stages to abrupt outbreaks, while a cooperative manipulation detection module explicitly identifies coordinated interactions formed by organized user groups and automated accounts. Extensive experiments on real-world multi-platform live-streaming e-commerce datasets demonstrate that MM-FGDNet consistently outperforms representative baseline methods, achieving an AUC of 0.927 and an F1 score of 0.847, with precision and recall reaching 0.861 and 0.834, respectively, while substantially reducing false alarm rates. Moreover, the proposed framework attains an Early Detection Score of 0.689. This metric serves as a critical benchmark for operational viability, quantifying the system's capacity to shift platform governance from passive remediation to proactive prevention. It confirms the reliable identification of the ``weak-signal'' stage---rigorously defined as the incipient phase where subtle, synchronized deviations in interaction rhythms manifest prior to traffic inflation outbreaks---thereby providing the necessary time window for preemptive intervention against coordinated manipulation. Ablation studies further validate the independent contributions of each core module, and cross-domain generalization experiments confirm stable performance across new streamers, new product categories, and new platforms. Overall, MM-FGDNet provides an effective and scalable data-driven artificial intelligence solution for early detection of coordinated abnormal behaviors in live-streaming systems.}}


\begin{document}

\section{{Introduction} %MDPI: This is a reminder: The companies/manufacturers of chemicals & reagents, devices, instruments, commercial cell lines/samples/materials should be indicated together with their city (states abbreviation is required for USA and Canada) and country in their first appearance. Please add necessary information for those items in text. For all the software used, please add the version number of them. Please check full text.
}
\label{sec: introduction}

With the rapid development of the digital economy and platform-based business models, live-streaming e-commerce, short-video marketing, and social media content distribution have become deeply integrated, gradually forming a digital marketing ecosystem driven by user interaction behaviors \cite{rachmad2024evolution}. Within this ecosystem, metrics such as watch duration, comment content, likes, and tipping behaviors not only directly influence product conversion rates but also serve as critical signals for platform recommendation mechanisms and traffic allocation strategies \cite{yang2024study}. However, alongside the continuous amplification of commercial value, a wide range of profit-driven abnormal marketing and fraudulent behaviors have emerged with increasing frequency, including fake tipping, order brushing, traffic inflation, bot-generated comments, coordinated opinion manipulation by organized groups, fabricated follower growth, and emotional manipulation \cite{fan2024regulation}. For instance, organized groups often utilize high-frequency fake tipping to trigger algorithmic promotion mechanisms, leading to artificial traffic inflation that unfairly displaces legitimate content. By creating such artificial popularity, these behaviors mislead consumer decision-making, undermine the fairness of platform ecosystems, and significantly increase the difficulty of platform governance and regulatory enforcement \cite{chen2023bright}. Unlike traditional e-commerce scenarios, live-streaming e-commerce is characterized by strong temporal dynamics, high interactivity, and the coexistence of multiple information modalities \cite{wang2021live}. Crucially, abnormal marketing behaviors manifest across these heterogeneous modes: a sudden surge in transaction volume (behavioral modality) may contradict the flat emotional trajectory of the streamer (audio/visual modality), or a flood of bot-generated comments (textual modality) may appear semantically disjointed from the current video context. These behaviors are rarely isolated events; instead, they tend to emerge gradually through coordinated changes in streamer behaviors, comment sentiment, transaction rhythms, and user interaction networks \cite{wu2023malicious}. As a result, single-modality or static features are insufficient to comprehensively capture the evolutionary process of fraudulent behaviors \cite{gan2024review}. Therefore, the development of an intelligent detection framework capable of integrating multimodal information, modeling temporal evolution, and identifying cooperative manipulation patterns is of significant theoretical value and practical importance for safeguarding platform content ecosystems, protecting consumer rights, and improving regulatory efficiency.

Early studies on marketing fraud detection primarily relied on statistical modeling and manual feature engineering, such as threshold-based abnormal traffic detection, time-series statistical feature analysis, and anomaly scoring methods based on user behavior frequency \cite{zhang2021hoba}. Although these methods demonstrate certain effectiveness in scenarios with clear rules and stable patterns, they heavily depend on prior experience and struggle to adapt to complex and rapidly evolving live-streaming marketing environments~\cite{baesens2021data}. Subsequently, machine learning models \cite{zhang2021new}, including logistic regression, support vector machines, and random forests, were introduced to perform fraud classification based on manually designed feature vectors \cite{nyasala2025predictive}. However, such approaches typically assume relatively stable data distributions and require high-quality labeled data \cite{li2023privacy, olushola2024fraud}. In real-world live-streaming scenarios, abnormal marketing behaviors are often characterized by extreme class imbalance, rapid evolution of fraud patterns, and high annotation costs, which significantly limits model generalization performance \cite{wang2024analysis}. In addition, some studies have employed graph-based models or social network analysis techniques to characterize user interaction relationships for identifying order-brushing groups or abnormal user communities. Nevertheless, these methods predominantly focus on structural information while overlooking the temporal evolution of behaviors and cross-modal semantic consistency, making it difficult to explain the coordinated manifestations of complex fraudulent behaviors across content, emotion, and transaction dimensions \cite{zhang2025identification}. Overall, conventional approaches generally suffer from the following limitations: $(1)$ reliance on single-modality or static features, resulting in limited sensitivity to cross-modal abnormal signals; $(2)$ insufficient capability to detect small-sample, weak-signal, and covert fraud patterns; and $(3)$ poor adaptability to distribution shifts across platforms, streamer types, and product categories.

In recent years, deep learning has achieved remarkable progress in computer vision, natural language processing, and speech analysis, providing new technical foundations for fraud detection in complex marketing scenarios \cite{malik2025future}. Several studies have explored the use of convolutional neural networks or transformer architectures for video content analysis to identify abnormal behaviors or emotional variations, while text models based on sentiment analysis and semantic representation learning have been applied to detect abnormal comments and coordinated spam content \cite{kokab2022transformer}. Moreover, recurrent neural networks and transformer-based models have demonstrated strong capabilities in modeling temporal dependencies within user behavior sequences \cite{hasan2023natural}. Building upon these advances, multimodal deep learning approaches have gradually attracted attention by integrating visual, textual, audio, and behavioral sequence information to enhance the perception of complex abnormal patterns \cite{ayetiran2024inter}. Despite these advances, significant challenges remain. First, substantial discrepancies in temporal granularity and semantic representation exist across different modalities, rendering simple concatenation or attention-based fusion insufficient for effective alignment \cite{xiao2022survey}. Second, most existing models rely on large-scale labeled data for supervised training, leading to pronounced performance degradation in scenarios with scarce fraud samples or noisy annotations \cite{li2024you, song2022learning}. Third, the organizational structure of cooperative behaviors is often neglected, limiting the ability to identify artificially generated popularity driven by coordinated bot clusters or organized groups \cite{chen2024comment}. Consequently, achieving cross-modal temporal alignment under weak supervision, modeling the dynamic evolution of abnormal behaviors, and explicitly characterizing cooperative manipulation patterns remain critical open problems. Khodabandehlou et al. \cite{khodabandehlou2024fifraud} proposed FiFrauD, an unsupervised financial fraud detection framework that performs temporal representation learning on dynamic transaction graph streams by jointly modeling node structural features and behavioral evolution patterns. Under unsupervised settings, effective identification of financial fraud was achieved, with significant improvements over traditional static graph models and statistical anomaly detection methods on multiple real-world financial datasets, particularly in terms of AUC, detection accuracy, early risk identification capability, and generalization performance in dynamic graph scenarios. Chen et al. \cite{chen2022analysis} proposed an opinion evolution analysis framework for live-streaming e-commerce regulation policies based on online comment mining, in which LDA topic modeling and sentiment analysis were employed to reveal dynamic changes in public sentiment and attention focus following policy implementation, enabling quantitative characterization of policy effects and behavioral patterns. Shehnepoor et al. \cite{shehnepoor2021hin} introduced the HIN-RNN heterogeneous information network representation learning framework, which constructs meta-path-guided node sequences and leverages recurrent neural networks to automatically learn high-order cooperative behavior patterns among fraudsters without manual feature engineering, significantly improving AUC and recall for fraud group detection on real-world datasets. Yu et al. \cite{yu2024temporal} proposed the TGFDN group fraud detection framework, which incorporates a Temporal Group Dynamics Analyzer to model the temporal evolution of group behaviors in transaction networks, achieving substantial improvements in accuracy and recall for group fraud identification on large-scale e-commerce and bitcoin datasets. Li et al. \cite{li2021live} developed the LIFE heterogeneous graph neural network framework, which integrates multiple entity relationships among users, streamers, and products and introduces a label propagation mechanism, leading to significant gains in abnormal transaction detection accuracy and recall in live-streaming e-commerce fraud detection tasks.

To address the aforementioned challenges, a multimodal fraud behavior detection framework for live-streaming e-commerce, termed MM-FGDNet, is proposed, with the main contributions summarized as follows:

\begin{enumerate}

\item A multimodal abnormal marketing dataset encompassing video, text, audio, and user behavior sequences is constructed, together with data synchronization and quality calibration strategies to enhance cross-modal consistency and reliability.

\item A cross-modal temporal alignment module is proposed, which integrates dynamic time alignment and semantic attention mechanisms to achieve effective fusion of heterogeneous modalities within a unified temporal semantic space.

\item A transformer-based temporal anomaly modeling module is designed to capture early weak signals and abrupt abnormal patterns of fraudulent behaviors, thereby enhancing the detection capability for covert marketing fraud.

\item A cooperative behavior detection module is introduced, in which graph neural networks are employed to model user interaction structures, enabling the identification of organized groups such as paid posters and bot clusters, along with interpretable abnormal subgraph analysis.

\item Diffusion models and self-supervised learning strategies are incorporated to improve model generalization performance and cross-scenario transferability under weakly labeled and small-sample conditions.

\end{enumerate}

The proposed MM-FGDNet systematically characterizes abnormal marketing behaviors in live-streaming e-commerce from three complementary perspectives: multimodal temporal alignment, temporal anomaly evolution modeling, and cooperative behavior identification. Compared with existing approaches, the proposed framework is capable of capturing cross-modal weak signals and identifying complex fraud patterns organized through multi-user coordination, demonstrating significant advantages in early risk detection and cross-platform generalization, and providing a practically valuable intelligent detection solution for digital marketing regulation and platform governance.

\section{Related Work}

\subsection{Multimodal Temporal Representation Learning}

The core objective of multimodal temporal representation learning lies in jointly modeling dynamic information from different perceptual channels and characterizing their complementary relationships within a unified temporal semantic space \cite{li2024improving}. In live-streaming e-commerce and digital marketing scenarios, visual content, speech signals, textual comments, and user interaction behaviors collectively constitute a complex, time-evolving marketing context \cite{wan2024optimal}. Existing studies have demonstrated that relying on a single modality is often insufficient for capturing rapidly changing behavioral nuances, whereas multimodal fusion can substantially enhance the representation of complex temporal semantics \cite{xu2024memf}. In the visual modality, approaches such as temporal convolutional networks or transformer architectures are commonly deployed to characterize streamers’ body movements, facial expressions, and scene transitions, thereby capturing continuous behavioral cues related to user engagement \cite{porras2025face}. With the introduction of self-attention mechanisms, these models can effectively capture long-range temporal dependencies, enabling a more comprehensive understanding of behavioral evolution throughout live-streaming sessions \cite{zhang2025forecasting}. However, such approaches typically prioritize content representation and rarely model the dynamic, non-linear interactions between content evolution and user feedback \cite{zhang2025product}.

For the speech and audio modality, sequential modeling techniques for emotion recognition and prosody analysis have been widely adopted to characterize streamers’ emotional states and expressive styles over time \cite{li2023oh}. Deep neural networks based on acoustic features can reflect the rhythm of live-streaming and instantaneous emotional fluctuations, providing potential temporal cues for abnormal behavior analysis \cite{luo2021study}. Nevertheless, the causal relationship between vocal emotion and marketing outcomes is highly context-dependent, and audio features alone are often insufficient to yield stable risk judgments \cite{lin2023review}. Similarly, research on the textual modality has shifted from static sentiment analysis to dynamic modeling of comments and bullet screens \cite{xiong2024exploring}. Methods based on pretrained language models are capable of capturing emotional tendencies, topic shifts, and potential manipulation traces in text streams \cite{kwon2024sentiment}. However, in live-streaming e-commerce scenarios, the text data is characterized by high noise, short length, and extreme repetition, making it difficult for standard sequential text models to distinguish genuine high-frequency interactions from organized comment brushing behaviors \cite{zhou2025study}.

To bridge these gaps, cross-modal representation learning aims to enable information interaction across modalities through shared embedding spaces or attention mechanisms~\cite{2022cross}. Representative approaches include fusion models based on alignment learning and contrastive learning, whose objective is to establish semantic consistency among different modalities \cite{yuan2025cross}. Despite these advancements, existing multimodal temporal paradigms face significant limitations when addressing organized group-level manipulation in real-time. Most current approaches focus on the temporal evolution of individual users or isolated content semantics, lacking the capability to explicitly model the topological structure of coordinated interactions among multiple accounts. Consequently, these paradigms struggle to distinguish between genuine organic popularity and synchronized artificial traffic generated by bot clusters or ``water armies,'' rendering them insufficient for the timely detection and mitigation of collaborative fraud patterns in highly dynamic live-streaming environments.

\subsection{Graph-Based Relational Structure Modeling}

While fundamental anomaly detection relies on identifying deviations from normal patterns \cite{popoola2023big}, traditional statistical and time-series approaches---ranging from distribution analysis \cite{khodabandehlou2024abnormal} to transformer-based temporal modeling \cite{thundiyil2023transformers}---predominantly focus on individual behavioral sequences. These methods exhibit advantages in detecting sudden trend changes but struggle to identify abnormal patterns formed through multi-user coordination \cite{hamidi2024approach}. To address this limitation, graph-based relational structure modeling has been widely adopted to characterize the topological dependencies among users. By constructing user--user or user--item interaction networks, these approaches leverage graph neural networks to learn high-order relational information, showing strong potential in detecting group-based order brushing and cooperative fraud structures that are invisible to independent sequence analysis \cite{ren2023dynamic}.

{However, a critical bottleneck remains in the current landscape: most existing graph models treat user interactions as static topological snapshots or overlook the fine-grained temporal evolution of node behaviors alongside multimodal content \cite{wen2022trend}. By separating structural inference from temporal dynamics and content semantics, existing paradigms fail to capture the continuous, dynamic formation process of organized group-level manipulation in real-time. Consequently, these methods are often unable to distinguish between organic community growth and the rapid, synchronized deployment of ``water armies'' or bot clusters during live-streaming events, leading to a lag in risk response.}

\subsection{{Generative and Weakly Supervised Learning Paradigms}}

{In real-world marketing scenarios, fraud samples are typically scarce and costly to annotate, posing severe challenges to traditional supervised learning methods \cite{walauskis2025unsupervised}. To address this, the central idea of weakly supervised learning is to reduce dependence on manually labeled data by exploiting intrinsic data structures or generating auxiliary samples to enhance model learning capability \cite{njima2022dnn}. Self-supervised learning paradigms enable models to learn general representations from unlabeled data by designing surrogate prediction tasks, thereby providing effective initialization for downstream tasks \cite{lee2021predicting}. In the context of fraud detection, self-supervised approaches are employed to model the temporal regularities of normal behaviors, such that abnormal behaviors naturally deviate from the normal distribution in the representation space \cite{liu2021self}. Furthermore, contrastive learning strengthens sensitivity to subtle differences by constructing positive and negative sample pairs, demonstrating promising performance in anomaly detection and user behavior modeling \cite{liu2021anomaly}.}

{Complementing these approaches, generative models provide an alternative solution to the few-shot problem. Methods based on generative adversarial networks can synthesize limited fraud samples to alleviate class imbalance, while diffusion models progressively model data distributions and exhibit stronger stability in generating high-quality temporal samples \cite{ghaleb2023ensemble}. Such generative approaches are beneficial for enriching the representational space of abnormal patterns when fraudulent interaction data are scarce \cite{lin2024diffusion}. Meanwhile, few-shot classification methods aim to achieve effective classification under extremely limited labeled samples through metric learning or prototypical learning \cite{gharoun2024meta}. However, in complex marketing scenarios, fraudulent behaviors are often highly diverse and covert, rendering it difficult for standalone few-shot classification methods to cover all potential patterns \cite{jin2025analysis}. More critically, while these paradigms effectively address data scarcity, they typically focus on individual sample augmentation or static distribution learning. They lack the mechanism to explicitly model the dynamic topological evolution of organized group-level manipulation, and thus fail to capture the real-time synchronized signals characteristic of collaborative fraud in live-streaming environments.}

\section{Materials and Method}
\label{sec: materials and method}
\subsection{Data Collection}

The data collection process in this study was conducted around real-world live-streaming e-commerce and digital marketing scenarios, with the objective of systematically characterizing the manifestations of abnormal marketing and fraudulent behaviors in multimodal spaces. The data were primarily collected from publicly accessible live-streaming rooms and product promotion scenarios on multiple mainstream live e-commerce and short-video platforms, with a continuous collection period spanning six months. This temporal coverage included different time intervals, streamer types, and product categories to reduce temporal bias and scenario-specific bias. The collected data encompassed multiple heterogeneous modalities, including live-streaming video streams, streamer speech signals, user comments and barrage text, tipping and transaction logs, and user interaction behavior sequences, as shown in Table~\ref{tab:data}. All data were obtained through publicly available platform interfaces, compliant crawling tools, and real-time stream monitoring, without involving any user private information.

For the visual modality, live-streaming video data were collected in raw stream format, and key frames were extracted at fixed temporal intervals to balance temporal completeness and computational efficiency. Video resolution and frame rate were preserved according to the original platform settings to ensure the authenticity of streamer facial expressions, body movements, and scene variations. Audio data were synchronously collected with video streams and processed using a unified sampling rate, enabling the extraction of vocal emotion, prosodic variations, and abnormal expressive patterns. Textual modality data included live comment streams and barrage streams, for which timestamps, user identifiers, and textual content were fully retained to support subsequent temporal analysis and cooperative behavior modeling. Behavioral and transaction data covered interaction events such as likes, follows, shares, tipping, and purchases, with millisecond-level timestamps, behavior types, and associated targets recorded to characterize behavior intensity, frequency variations, and bursty anomalies.

To support cross-modal analysis, all modalities were synchronously aligned through a unified timestamp mechanism, thereby constructing a consistent temporal index across data sources. During data collection, both normal marketing scenarios and abnormal marketing events confirmed through a combination of manual inspection and rule-based verification were included, covering typical patterns such as concentrated bursts of fake tipping, highly homogeneous comment content, and coordinated interactions among user groups. The resulting dataset exhibits strong representativeness in terms of temporal span, modal coverage, and behavioral complexity, providing a solid data foundation for subsequent multimodal alignment, temporal anomaly modeling, and cooperative behavior detection.

\begin{table}[H]
%\centering
%\small
\caption{Overview of the multimodal fraud detection dataset.}
\label{tab:data}

\begin{adjustwidth}{-\extralength}{0cm}
\centering %% If there is a figure in wide page, please release command \centering, for Table, ``\textwidth" should be ``\fulllength"
\begin{tabularx}{\fulllength}{lcCC}
\toprule
\textbf{Data Modality} & \textbf{Source} & \textbf{Quantity} & \textbf{Granularity} \\
\midrule
Live-stream video frames & Live e-commerce platforms & 3.2 M frames & Frame-level \\
Audio segments & Live-stream audio tracks & 18,400 clips & Segment-level \\
Comments \& barrage text & Live comment streams & 12.6 M entries & Message-level \\
Transaction records & Tipping and order logs & 1.8 M events & Event-level \\
User interaction sequences & Likes, follows, shares & 9.4 M actions & Action-level \\
\bottomrule
\end{tabularx}
\end{adjustwidth}
\end{table}

\subsection{Data Preprocessing and Augmentation}

In live-streaming e-commerce scenarios, multimodal data are characterized by heterogeneous sources, high noise levels, and inconsistent temporal granularity. Directly feeding raw data into learning models often leads to disordered feature distributions and the masking of abnormal patterns. Therefore, systematic preprocessing and augmentation of text, video, and user behavior sequences prior to modeling are essential for improving robustness and generalization performance. In this study, unified processing is conducted from multiple perspectives, including multimodal signal denoising, explicit modeling of semantic and emotional information, temporal behavior normalization, and synthesis of rare fraud patterns, thereby laying a solid foundation for subsequent cross-modal alignment and anomaly detection. For the textual modality, live-streaming comments and bullet-screen messages typically exhibit extremely short length, strong colloquiality, high redundancy, and frequent noise tokens. The primary objective of text preprocessing is to reduce the interference of irrelevant noise on semantic representations while preserving emotional and semantic cues indicative of abnormal marketing behaviors. Specifically, raw comment text is first processed through a normalization mapping function that converts emojis, abbreviations, and non-standard characters into unified semantic units. Let the original text sequence be denoted as $T=\{w_1,w_2,\dots,w_n\}$, and the normalized text representation can be expressed as
\begin{equation}
\tilde{T} = \phi(T),
\end{equation}
where $\phi(\cdot)$ denotes the text cleaning and standardization mapping operator. On this basis, a contextual encoding model is employed to extract semantic representations $E_t$, while an emotion analysis module is introduced to model the emotional states of the text. Emotional labels can be regarded as projections of the text in an emotional space, and their computation can be formulated as
\begin{equation}
s_t = f_{\text{emo}}(E_t),
\end{equation}
where $f_{\text{emo}}(\cdot)$ denotes the emotion prediction function, and $s_t$ represents the emotion score or emotion vector at the corresponding time step. Through joint modeling of semantic and emotional representations, the textual modality not only reflects comment content itself but also captures potential fraud signals such as emotional manipulation and abnormal opinion guidance.

In the video modality, live-streaming frames are often affected by illumination variations, compression artifacts, and motion blur, which may interfere with the analysis of streamer behaviors and scene changes. To address this issue, denoising and quality correction are first applied to video frame sequences. Let the original video frame at time $t$ be denoted as $I_t$, and the enhanced frame after denoising and illumination correction is given~by
\begin{equation}
\tilde{I}_t = \psi(I_t),
\end{equation}
where $\psi(\cdot)$ denotes the video enhancement operator. As live-streaming sessions typically last for extended durations and exhibit high redundancy across consecutive frames, a key-frame selection strategy is further adopted to retain frames that are most discriminative for behavioral changes. Key-frame selection is achieved by measuring inter-frame feature differences, where the difference between adjacent frames is defined as
\begin{equation}
d_t = |F(I_t) - F(I_{t-1})|_2,
\end{equation}
with $F(\cdot)$ denoting the visual feature extraction function. When $d_t$ exceeds a predefined threshold $\delta$, the corresponding frame is retained as a key frame. Through these procedures, the video modality preserves essential behavioral information while significantly reducing redundant noise, which is beneficial for subsequent cross-modal temporal alignment.

User behavior sequences constitute a critical data source for characterizing abnormal marketing behaviors and can be viewed as multivariate time series containing events such as likes, comments, tipping, and following. Given the strong randomness and burstiness of user behaviors, directly using raw sequences may cause models to overemphasize short-term fluctuations. Accordingly, smoothing and differencing operations are applied to behavior sequences to highlight trend changes and abnormal bursts. Let the original behavior sequence be denoted as $x_t$, and its smoothed representation is defined as
\begin{equation}
\bar{x}_t = \alpha x_t + (1-\alpha)\bar{x}_{t-1},
\end{equation}
where $\alpha \in (0,1)$ is the smoothing coefficient. To characterize behavior change rates, a first-order difference is introduced as
\begin{equation}
\Delta x_t = x_t - x_{t-1}.
\end{equation}

{In} %MDPI: Newly added indent format. Please confirm and check full text.
%Authors: We confirmed and checked.
 addition, certain abnormal marketing behaviors exhibit clear periodic patterns, such as concentrated traffic boosting within fixed time intervals. To capture such characteristics, periodic pattern analysis is incorporated by mapping behavior sequences into the frequency domain via the Fourier transform, whose spectral representation is given by
\begin{equation}
X(f) = \sum_{t=1}^{T} x_t e^{-j2\pi ft}.
\end{equation}

This transformation enables the identification of periodic anomalies hidden within time-series data, allowing the model to detect not only instantaneous abnormalities but also long-term and regular fraudulent behaviors. As genuine fraud samples typically account for a very small proportion of the overall data, training models solely on original data often leads to severe class imbalance and overfitting. To mitigate this issue, generative data augmentation strategies are introduced to synthesize rare abnormal samples by learning latent distributions of fraudulent behaviors. Within the generative adversarial network framework, a generator $G$ takes random noise $z$ as input and outputs a synthetic sequence $\hat{x}$, and its objective function can be expressed as
\begin{equation}
\min_G \max_D \; \mathbb{E}_{x \sim p_{\text{data}}}[\log D(x)] + \mathbb{E}_{z \sim p(z)}[\log(1-D(G(z)))],
\end{equation}
where $D$ denotes the discriminator. To further improve the temporal continuity and distributional stability of generated sequences, diffusion models are additionally employed to model abnormal behaviors. Diffusion models approximate real data distributions by gradually injecting noise into data and learning the reverse denoising process. The forward diffusion process can be formulated as
\begin{equation}
q(x_t|x_{t-1}) = \mathcal{N}(\sqrt{1-\beta_t}x_{t-1}, \beta_t I),
\end{equation}
{where $\beta_t$ denotes the noise scheduling parameter. By learning the reverse process $p_\theta(x_{t-1}|x_t)$, the model is capable of generating temporal segments that closely resemble real fraudulent behaviors. Specifically, this generative capability is leveraged to synthesize rare coordinated burst sequences, effectively alleviating the scarcity of organized attack samples and enhancing model robustness against high-intensity manipulation. Through the aforementioned multi-level preprocessing and augmentation strategies, textual, video, and behavioral sequences are mapped into representation spaces with lower noise, clearer semantics, and more balanced distributions prior to model input. This not only improves the accuracy of cross-modal alignment but also substantially enhances the capability of the model to identify covert fraud patterns under weak supervision, providing a solid data foundation for subsequent temporal anomaly modeling and cooperative behavior detection in MM-FGDNet.}

\subsection{Proposed Method}
\subsubsection{Overall}

A unified end-to-end modeling pipeline is adopted in the proposed multimodal fraud detection framework MM-FGDNet, where preprocessed and augmented multimodal temporal data are jointly modeled to systematically characterize the generation and evolution mechanisms of abnormal marketing behaviors from both temporal and group-structural perspectives. The overall framework is initiated with cross-modal temporal semantic alignment, through which feature representations derived from video, text, audio, and user behavior sequences are projected into a unified temporal semantic space, thereby eliminating discrepancies in sampling frequency, temporal granularity, and response latency across different modalities. Specifically, features from each modality are first fed into modality-specific encoders to obtain intra-modal temporal representations, which are then flexibly mapped through a dynamic temporal alignment mechanism, enabling consistent temporal correspondence of the same latent behavioral event across different modalities. On this basis, a semantic attention fusion mechanism is employed to adaptively integrate the aligned multimodal features, yielding a discriminative and temporally comparable multimodal joint representation at each time step.

After cross-modal alignment, the multimodal temporal representations are further processed by the temporal anomaly modeling module for deep sequential analysis. By leveraging self-attention structures, long-range dependencies among behavioral signals over extended time spans are explicitly captured, allowing early weak signals and subsequent abnormal bursts to be jointly perceived. Through multi-scale modeling of behavioral variations, both short-term abrupt anomalies and long-term covert fraud patterns emerging through gradual evolution can be effectively identified. The output of this module consists of temporally continuous risk representations, which provide fine-grained temporal characterization for downstream risk assessment. Meanwhile, user-level interaction behaviors are abstracted into dynamic graph structures and are simultaneously fed into the cooperative manipulation detection module. Graph neural networks are utilized to model user interaction topologies, enabling the propagation and aggregation of individual behavioral information within group structures and explicitly characterizing cooperative relationships among multiple users. The group-level features learned from graph modeling are jointly analyzed with the temporal risk representations produced by the temporal anomaly modeling module in a high-level semantic space, so as to determine whether detected anomalies originate from individual behavioral deviations or coordinated group manipulation. Through subgraph-level anomaly identification and cross-modal consistency verification, potential bot clusters or organized manipulation groups can be localized, and their contributions to abnormal popularity formation can be interpreted. Finally, outputs from the temporal anomaly modeling module and the cooperative manipulation detection module are fused at the decision layer to jointly produce time-step-level fraud probabilities, abnormal behavior category predictions, and comprehensive risk scores. Through this bottom-up modeling process that progresses from temporal dynamics to group structures, MM-FGDNet enables systematic identification of complex abnormal marketing behaviors in live-streaming e-commerce scenarios and provides a unified and effective technical framework for early risk warning and interpretable regulation.

\subsubsection{Cross-Modal Temporal Alignment}

\begin{figure}[H]
\includegraphics[width=\linewidth]{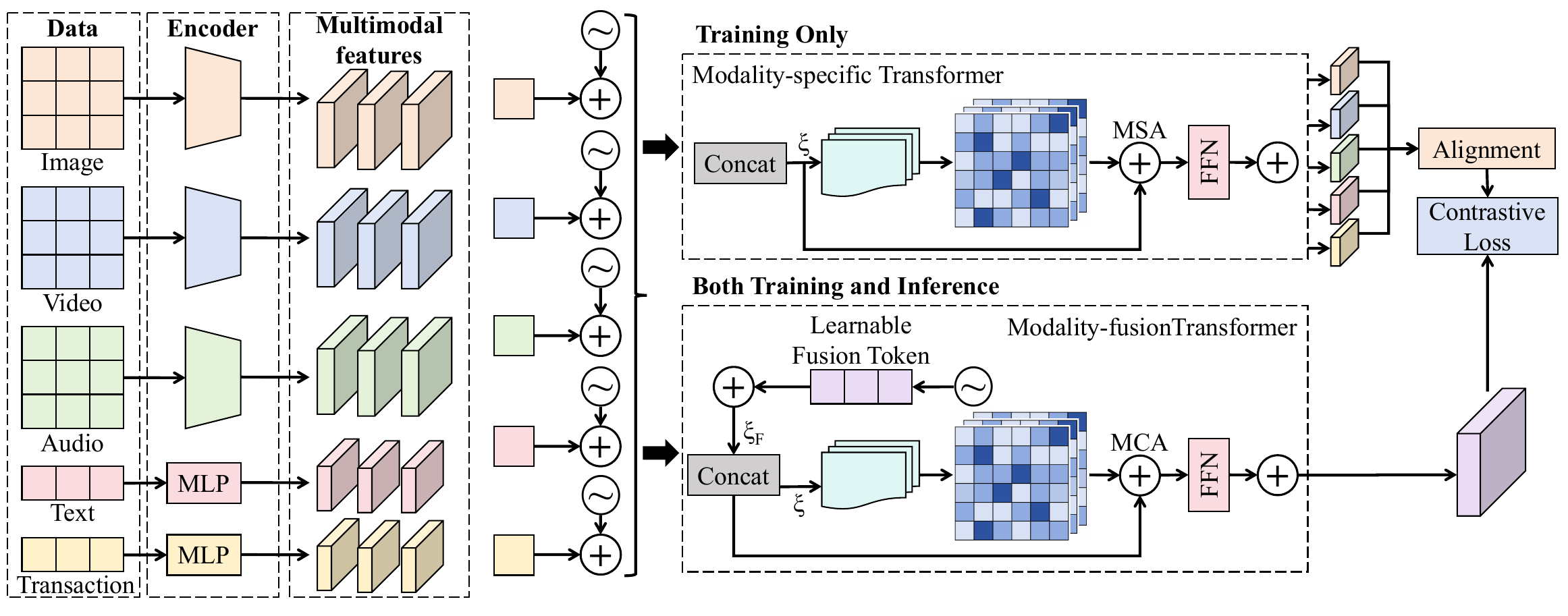}
\caption{{Illustration} %MDPI: 1.Please confirm if the color needs to be explained. 2.We moved Figure 1 to reduce the blank. Please confirm.
%Authors: 1. We have confirmed that no further explanation is required. 2. We confirmed.
 of the cross-modal temporal alignment module.}
\label{fig: stride}
\end{figure}

The cross-modal temporal alignment module constitutes a core component of MM-FGDNet, aiming to establish unified and comparable temporal semantic representations across multimodal signals with heterogeneous temporal granularities, semantic response delays, and noise characteristics, such that responses of the same latent marketing behavior in video, text, audio, and transaction modalities can be consistently aligned along the temporal axis. As shown in Figure~\ref{fig: stride}, preprocessed multimodal features are first fed into modality-specific encoders to obtain intra-modal temporal feature sequences. Lightweight temporal encoding networks are employed for visual and audio modalities to preserve continuous dynamic information, while text and transaction modalities are projected into the same latent space dimension as other modalities via multilayer perceptrons, achieving preliminary alignment at the feature dimension level. Let the encoded representation of the $m$-th modality at time step $t$ be denoted as $h_t^{(m)} \in \mathbb{R}^d$, where $d$ represents the unified latent space dimension.

To address asynchronous responses across modalities along the temporal axis, a dynamic temporal alignment mechanism is introduced to learn flexible temporal mappings for modality-specific features. Instead of relying on fixed windows or hard alignment strategies, this mechanism establishes soft associations between neighboring time steps via learnable alignment weight functions, allowing the model to automatically capture realistic behavioral response patterns, such as streamer actions preceding comment sentiment changes or abnormal tipping lagging behind content stimulation. Through this process, original modality-specific temporal sequences are remapped to a shared temporal index $t'$, yielding aligned feature representations $\tilde{h}_{t'}^{(m)}$ and effectively eliminating structural biases caused by sampling frequency differences and response delays. After temporal alignment, a semantic attention fusion structure is further employed to jointly model multimodal information. At each aligned time step $t'$, features from different modalities are regarded as multi-view expressions of the same behavioral event, and modality weights are dynamically assigned via attention mechanisms, such that modalities contributing more to anomaly discrimination receive higher importance. This strategy not only avoids noise amplification induced by naive feature concatenation but also enables adaptive modulation of modality importance across different stages, for example, relying more on textual and behavioral signals during early anomaly stages and emphasizing visual and transaction modalities during abnormal bursts. The fused multimodal representation can be expressed as $z_{t'}=\sum_m \alpha_{t'}^{(m)} \tilde{h}_{t'}^{(m)}$, where $\alpha_{t'}^{(m)}$ denotes the modality weight learned by the attention function.
\vspace{-4pt}

To further enhance cross-modal semantic consistency, a cross-modal contrastive constraint is incorporated during training, encouraging multimodal representations corresponding to the same temporal semantics to remain close in the latent space, while representations from unrelated time steps or inconsistent behaviors are pushed apart. This design explicitly suppresses modality drift induced by noise or incidental events and strengthens the focus on genuine behavior-driven signals. From a mathematical perspective, this constraint effectively minimizes distances among modality representations at the same $t'$ while maximizing discriminability across different time steps or samples, thereby improving both the discriminative power and stability of aligned representations. Combined with the shared fusion Transformer employed during both training and inference, this module significantly enhances alignment quality and downstream anomaly detection performance while maintaining inference efficiency. Overall, through the synergistic design of intra-modal encoding, flexible temporal alignment, semantic attention fusion, and cross-modal consistency constraints, the cross-modal temporal alignment module effectively addresses asynchronous responses, semantic misalignment, and noise interference in multimodal live-streaming e-commerce data. As a result, highly consistent multimodal input representations in both temporal and semantic dimensions are provided for subsequent temporal anomaly modeling, substantially improving the model's sensitivity to weak signals and covert abnormal marketing behaviors.

\subsubsection{Temporal Fraud Pattern Modeling}

\begin{figure}[H]
\includegraphics[width=0.99\linewidth]{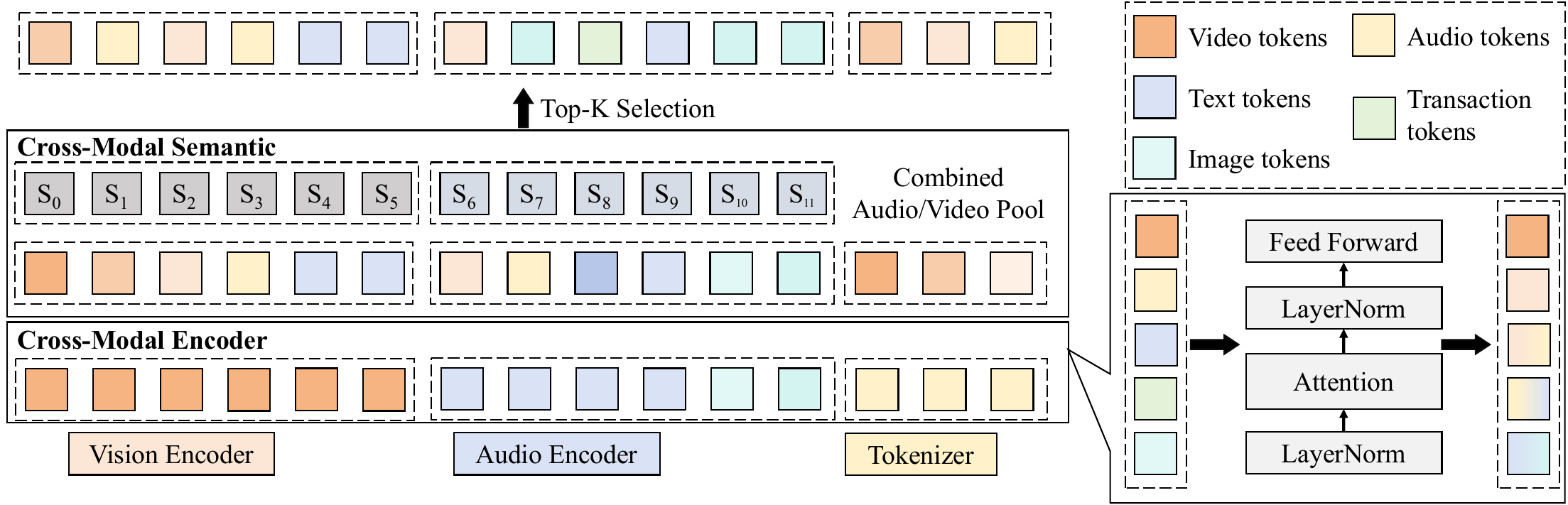}
\caption{This figure illustrates the architecture and information flow of the temporal fraud pattern modeling module.}
\label{fig: time}
\end{figure}

The temporal fraud pattern modeling module takes the unified temporal semantic representations produced by the cross-modal temporal alignment module as input. Let the aligned sequence be denoted as $X \in \mathbb{R}^{T \times d}$, where $T$ represents the length of the aligned temporal sequence, which can be regarded as the temporal height, and $d$ denotes the feature width, which is set to $256$ in the implemented configuration. To balance the requirements of long live-streaming sequences and computational efficiency, a cross-modal semantic compression strategy is first applied, in which highly redundant audio and video token pools are jointly selected and merged with text and transaction tokens to form a compact cross-modal temporal token sequence. The compressed length is denoted as $K$, which is typically set to $128$, ensuring that subsequent temporal modeling focuses on key segments with higher information density. Subsequently, a hierarchical temporal Transformer encoder is employed to model the evolution of abnormal behaviors, with the network depth set to $L = 6$ layers. Each layer contains $H = 8$ attention heads, where the dimension of each head is $d/H = 32$, and the feed-forward network dimension is set to $d_{ff} = 1024$. A pre-norm residual structure is adopted to stabilize training on long sequences. For the $\ell$-th layer, the attention computation is formulated as
\begin{equation}
Q^{(\ell)} = X^{(\ell-1)} W_Q^{(\ell)}, \quad
K^{(\ell)} = X^{(\ell-1)} W_K^{(\ell)}, \quad
V^{(\ell)} = X^{(\ell-1)} W_V^{(\ell)},
\end{equation}
\begin{equation}
\mathrm{Attn}^{(\ell)}(X^{(\ell-1)}) =
\mathrm{softmax}\!\left(\frac{Q^{(\ell)} {K^{(\ell)}}^\top}{\sqrt{d/H}} + M \right) V^{(\ell)},
\end{equation}
where $M$ denotes a causal or local-window mask. A hybrid masking strategy that combines local windows with a small number of global tokens is adopted, enabling the model to stably capture short-term bursts while establishing long-range dependencies through global tokens. To further enhance local--global pattern fusion, a lightweight temporal convolution branch is introduced in parallel after each attention output to amplify local abrupt variations. The convolution kernel size is set to $k = 3$, and dilation rates $r \in \{1,2,4\}$ are alternated across layers, allowing the effective receptive field to expand with depth. The fused representation is expressed as
\begin{equation}
U^{(\ell)} =
\mathrm{Attn}^{(\ell)}(X^{(\ell-1)}) +
\sum_{r \in \{1,2,4\}} \mathrm{Conv}_{k,r}\!\left(X^{(\ell-1)}\right),
\end{equation}
followed by a feed-forward transformation
\begin{equation}
X^{(\ell)} =
U^{(\ell)} +
\sigma\!\left(U^{(\ell)} W_1^{(\ell)}\right) W_2^{(\ell)},
\end{equation}
where $\sigma(\cdot)$ denotes the gelu activation function. To improve sensitivity to weak signals without excessive reliance on explicit labels, a set of implicit fraud pattern embeddings is introduced as a learnable prototype set $P = \{p_1, \dots, p_M\} \subset \mathbb{R}^d$, with $M = 16$. Prototype attention is used to map temporal representations to anomaly evidence strength:
\begin{equation}
a_t =
\max_{m \le M}
\frac{\exp\!\left(\tau \cdot \langle x_t, p_m \rangle\right)}
{\sum_{j=1}^{M} \exp\!\left(\tau \cdot \langle x_t, p_j \rangle\right)},
\end{equation}
{where $x_t$ denotes the final-layer representation at time step $t$, and $\tau$ is a temperature parameter. This design allows stable amplification of weak signals when fraud patterns form clusters in the latent space. Specifically, the optimization of $\tau$ is critical for balancing sensitivity and precision; by tuning $\tau$ to sharpen the attention distribution only when prototype similarity exceeds a confidence margin, the mechanism acts as a soft-thresholding filter. This ensures that genuine weak signals aligning with learned prototypes are amplified, while random noise with low similarity scores is suppressed, thereby preventing non-specific fluctuations from elevating the false alarm rate. Furthermore, to enable interpretable detection of abrupt anomalies, a temporal variation energy is defined by jointly considering adjacent state differences and prototype evidence:}
\begin{equation}
e_t = \|x_t - x_{t-1}\|_2^2 + \lambda a_t,
\end{equation}
which is used to produce time-step-level risk logits $\ell_t = w^\top x_t + b + \eta e_t$, and the final fraud probability is obtained as $\hat{y}_t = \sigma(\ell_t)$. A key advantage of combining this module with the cross-modal temporal alignment module lies in the fact that aligned multimodal responses are mapped into a shared temporal semantic space, such that evidence of the same latent event forms consistent correlation structures within $X$. Consequently, the attention matrix $Q K^\top$ becomes more block- or band-separable, reducing attention noise caused by semantic misalignment. Formally, if alignment errors randomly mismatch a proportion $\epsilon$ of truly correlated pairs, the expected signal-to-noise ratio of attention scores satisfies $\mathbb{E}[\Delta] \ge (1-\epsilon)\Delta_0 - \epsilon\Delta_1$, where $\Delta_0$ denotes the average margin of correctly matched pairs and $\Delta_1$ represents an upper bound on mismatched margins, implying that reducing $\epsilon$ linearly improves separability and weak-signal detection, as shown in Figure~\ref{fig: time}. 

Finally, the necessity of local--global fusion can be explained by the fact that when only local convolutions are used, the effective receptive field of a network with depth $L$ is at most $1 + 2 \sum_{\ell=1}^{L} r_\ell$, which is insufficient for long live-streaming sequences. By introducing global attention, any time step $t$ can establish a one-hop dependency with any other step $s$, reducing the upper bound of influence path length from $O(T)$ to $O(1)$, which can be characterized as
\begin{equation}
\forall t, s, \quad
\frac{\partial x_t^{(\ell)}}{\partial x_s^{(\ell-1)}} \neq 0 \ \text{{may hold through attention}%MDPI: Please confirm whether it needs to be moved to the main text.
%Authors: We've confirmed, it's not needed.
}.
\end{equation}

This ensures that early weak signals can influence subsequent decisions across long time spans, while the convolution branch remains sensitive to high-frequency components of abrupt spikes. Their complementarity enables both early trend recognition and instantaneous burst capture, which is consistent with the progressive evolution and sudden amplification characteristics of abnormal marketing behaviors in live-streaming e-commerce.

\subsubsection{Cooperative Manipulation Detection}

\begin{figure}[H]
\includegraphics[width=0.99\linewidth]{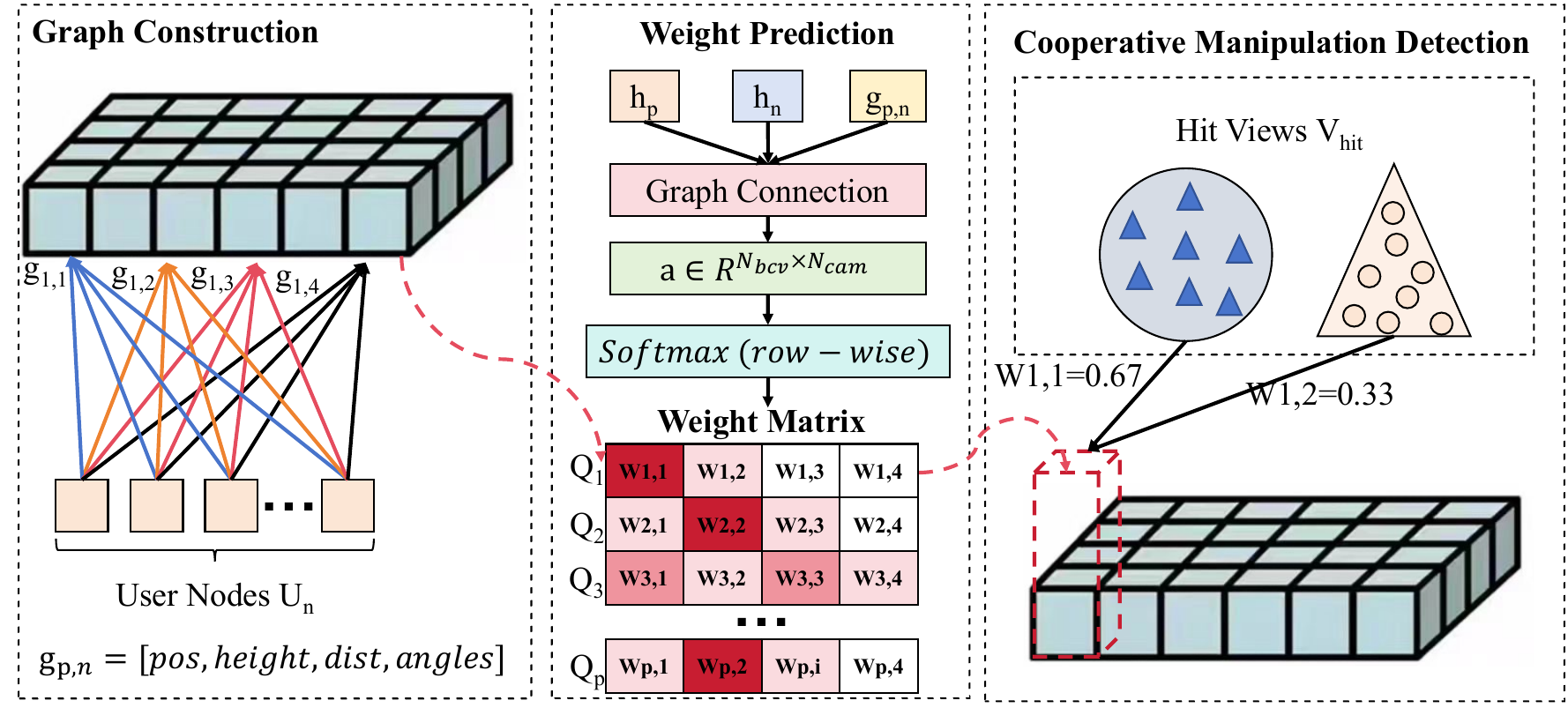}
\caption{{This} %MDPI: Please confirm whether the dashed arrow needs an explanation.
%Authors: We have confirmed that no further explanation is required.
 figure depicts the overall pipeline and structural design of the cooperative manipulation detection module.}
\label{fig: co}
\end{figure}

The cooperative manipulation detection module takes the user interaction graph as its core input and explicitly models the organizational structures of coordinated manipulation, such as bot clusters and paid commentator groups, which constitute a key fraud mechanism in live-streaming e-commerce. This module forms a complementary closed loop with the cross-modal temporal alignment and temporal fraud pattern modeling modules. As shown in Figure~\ref{fig: co}, within a live-streaming time window of length $T$, a dynamic graph $G = (V, E)$ is constructed, where $|V| = N$ denotes the number of user nodes, which can be regarded as the graph height, and each edge $e_{ij} \in E$ represents an interaction between users $i$ and $j$ within the window, such as comment replies, co-tipping to the same streamer, mutual mentions, or short-term synchronized likes. Edge weights are determined by interaction intensity and temporal proximity. The initial representation of each node is composed of multi-source behavioral statistics and aligned content evidence, denoted as $h_i^{(0)} \in \mathbb{R}^{d_g}$, where $d_g = 256$ is the graph feature width. To inject cross-modal consistency evidence, the time-step-level risk sequence output by the temporal anomaly module is aggregated along user events to obtain a user-level risk summary $r_i \in \mathbb{R}^{64}$, which is concatenated with structural features and linearly projected before entering the graph network, enabling graph representations to simultaneously encode individual temporal anomalies and group structural relations.

The graph encoder adopts a $L_g = 3$-layer graph attention network to enhance the representation of heterogeneous relations and cooperative behaviors. The output channel dimensions are set to $256\rightarrow256\rightarrow128$, with the number of attention heads $H_g = 4$ and each head having a dimension of $64$. For the $\ell$-th layer, message passing is formulated as
\begin{equation}
e_{ij}^{(\ell)} =
\mathrm{LeakyReLU}\!\left(
a^{(\ell)\top}
\big[
W^{(\ell)} h_i^{(\ell-1)} \ \Vert \
W^{(\ell)} h_j^{(\ell-1)} \ \Vert \
g_{ij}
\big]
\right),
\end{equation}
where $g_{ij}$ denotes the edge attribute vector, including time difference, interaction type distribution, and shared targets, which is linearly projected to aligned dimensions. Row-wise softmax normalization is applied to obtain interpretable neighbor contribution weights
\begin{equation}
\alpha_{ij}^{(\ell)} =
\frac{\exp(e_{ij}^{(\ell)})}
{\sum_{k \in \mathcal{N}(i)} \exp(e_{ik}^{(\ell)})},
\end{equation}
and node updates are performed as
\begin{equation}
h_i^{(\ell)} =
\sigma\!\left(
\sum_{j \in \mathcal{N}(i)}
\alpha_{ij}^{(\ell)} W^{(\ell)} h_j^{(\ell-1)}
\right),
\end{equation}
which is consistent with the ``GAT--row-wise softmax--weight matrix'' pipeline illustrated in the architecture, allowing higher attention weights to be assigned to the most likely cooperative accomplices and highlighting strong intra-group connections. For subgraph-level anomaly detection, a learnable set of group prototypes $C = \{c_1, \dots, c_K\} \subset \mathbb{R}^{128}$ is introduced, with $K = 8$, and soft clustering is used to compute node memberships to each prototype:
\begin{equation}
\pi_{ik} =
\frac{\exp\!\left(\kappa \cdot h_i^{(L_g)\top} c_k\right)}
{\sum_{u=1}^{K} \exp\!\left(\kappa \cdot h_i^{(L_g)\top} c_u\right)}.
\end{equation}

Based on this, subgraph risk is defined as a joint function of prototype consistency and edge-level cooperation strength:
\begin{equation}
s_{\text{sub}} =
\sum_{k=1}^{K}
\left\|
\sum_{i \in V} \pi_{ik} h_i^{(L_g)} - \mu_k
\right\|_2^2
+
\rho
\sum_{(i,j) \in E}
\sum_{k=1}^{K}
\pi_{ik} \pi_{jk} w_{ij},
\end{equation}
{where $\mu_k$ denotes the running mean of the $k$-th prototype center and $w_{ij}$ is the edge weight. The second term mathematically encourages nodes connected by high-strength edges to achieve higher consistency under the same prototype, which corresponds to the organizational characteristics of coordinated manipulation. In this context, the coefficient $\rho$ serves as a regularization parameter that calibrates the trade-off between feature consistency and structural density. Optimizing $\rho$ ensures that high interaction density alone---which may appear in legitimate fan communities---does not dominate the risk score. Instead, it enforces a dual constraint where subgraphs are only flagged if they simultaneously exhibit tight coordination and alignment with fraud prototypes, effectively amplifying covert organized signals without misclassifying benign group activities. For cross-modal consistency verification, user embeddings from content and behavior perspectives, denoted as $u_i^{c}$ and $u_i^{b}$, are constructed using aligned temporal semantics, and their discrepancy is used as evidence of pseudo-cooperation:}
\begin{equation}
\Delta_i =
1 -
\frac{u_i^{c\top} u_i^{b}}
{\|u_i^{c}\|_2 \|u_i^{b}\|_2}.
\end{equation}

When a large number of nodes simultaneously exhibit low content diversity but high behavioral synchrony, both $\Delta_i$ and $s_{\text{sub}}$ increase, facilitating the localization of bot clusters. The advantage of combining this module with the temporal anomaly modeling module lies in the complementary nature of evidence: the temporal module provides fine-grained information about when anomalies occur, while the graph module identifies who is cooperating. These signals are fused at the decision layer via learnable gating, yielding the final risk score $\hat{y} = \sigma(\theta_1 \bar{p} + \theta_2 s_{\text{sub}} + \theta_3 \bar{\Delta})$, where $\bar{p}$ denotes aggregated temporal risk and $\bar{\Delta}$ denotes aggregated consistency discrepancy. It can be shown that when intra-group edge weights $w_{ij}$ increase and group members become more concentrated in the prototype space, $s_{\text{sub}}$ increases monotonically, that is, $\partial s_{\text{sub}} / \partial w_{ij} \ge 0$, indicating that this metric provides an interpretable response to cooperation strength. Meanwhile, $\Delta_i$ is scale-invariant due to cosine discrepancy, allowing stable discrimination across different streamers and product categories, thereby significantly improving cross-scenario generalization and detection of covert cooperative fraud.

\section{Results and Discussion}
\label{sec: results and discussion}

\subsection{Experimental Configuration}

\subsubsection{Hardware and Software Platform}

All experiments were conducted on a unified high-performance computing platform to ensure stability and reproducibility during model training and evaluation. In terms of hardware configuration, the experimental server was equipped with multi-core high-performance CPUs and large-capacity memory to support parallel loading and preprocessing of large-scale multimodal data, and was further accelerated by high-performance GPUs for deep learning model training and inference. The GPU memory capacity was sufficient to accommodate simultaneous multimodal feature inputs and long-sequence modeling, thereby effectively avoiding gradient truncation or batch size reduction caused by memory limitations. A high-speed solid-state drive storage system was adopted to ensure efficient read and write operations for video frames, textual logs, and user behavior sequences, which collectively improved experimental efficiency and overall system throughput. Regarding the software environment, the experimental platform was built upon mainstream deep learning frameworks and deployed on a 64-bit Linux distribution to ensure robust support for GPU drivers and parallel computing libraries. Model implementation was completed using mature deep learning frameworks with automatic differentiation and distributed training mechanisms to enhance training stability. Common numerical computation and data analysis libraries were employed in multimodal data processing to support text processing, video decoding, feature extraction, and graph structure construction. In addition, random seeds were consistently controlled across experiments to reduce stochastic variation between different runs and to ensure strong reproducibility of experimental results.

{To strictly prevent future information leakage and simulate real-world streaming scenarios, the dataset was partitioned chronologically. Specifically, the first 70\% of the timeline was designated for training, the subsequent 10\% for validation, and the final 20\% for testing.} The training set was used for model parameter learning, the validation set for hyperparameter tuning and model selection, and the test set exclusively for final performance evaluation. To further enhance the reliability of experimental results, a five-fold cross-validation strategy was adopted on the entire dataset. Specifically, the dataset was divided into five mutually exclusive subsets, with one subset used as the test set in each experimental round and the remaining subsets used for training and validation. Final results were reported as the average performance over the five runs. During model training, an adaptive gradient-based optimization strategy was employed, with the initial learning rate set to $\alpha$ and gradually decayed according to a predefined schedule to balance convergence speed and training stability. The batch size was determined based on GPU memory capacity and model complexity to ensure stable gradient estimation. The hidden dimensions and numbers of attention heads in temporal models were tuned using the validation set to achieve an appropriate trade-off between performance and computational cost. Regularization weights and dropout probabilities were introduced to mitigate overfitting, and their values were determined through comparative experiments. All hyperparameters were kept consistent across cross-validation folds to ensure comparability of results.

\subsubsection{Baseline Models and Evaluation Metrics}

A diverse set of representative models was selected as baseline methods for comparison, including LSTM/GRU-based temporal models \cite{hochreiter1997long, cho2014properties}, transformer-based temporal models without multimodal fusion \cite{vaswani2017attention}, graph neural network based fraud detection models \cite{liu2020alleviating}, multimodal fusion models such as MMBT and CLIP-style fusion approaches \cite{kiela2019supervised,radford2021learning}, as well as commercial or publicly available anomaly detection models including isolation forest and one-class svm \cite{liu2008isolation, scholkopf2001estimating}. LSTM and GRU model user behavior sequences through gated recurrent mechanisms and are effective at capturing short-term temporal dependencies, benefiting from mature architectures and stable training behavior. Transformer-based temporal models rely on self-attention mechanisms to model global temporal dependencies and exhibit advantages in capturing long-range behavioral patterns. GNN-based fraud detection methods propagate node information over user interaction graphs to identify abnormal structures, demonstrating strong expressive power for detecting coordinated group behaviors. MMBT and CLIP-style fusion models jointly model textual and visual features to enhance cross-modal semantic alignment by leveraging complementary multimodal information. Isolation forest and one-class svm identify deviations by modeling the distribution of normal samples, requiring no large-scale labeled data and offering simple implementation, which makes them suitable as unsupervised baseline methods.

For experimental evaluation, a comprehensive set of metrics was adopted to assess discriminative capability, stability, and practical applicability in live-streaming e-commerce fraud detection tasks. These metrics include auc, precision, recall, and {f1} %MDPI: Please ensure all variables/values in the equation appear in the same format in the text (normal/italic/bold/subscript/superscript).
%Authors: We checked and confirmed.
 for overall classification performance; fraud detection rate (fdr) and false alarm rate (far) for characterizing risk identification effectiveness; and early detection score and cross-domain generalization score to reflect the capability of capturing early weak signals and generalizing across different scenarios. The mathematical definitions of the evaluation metrics are provided below, where the binary confusion matrix consists of true positives, false positives, true negatives, and false negatives:
\begin{equation}
\text{Precision} = \frac{TP}{TP+FP}, \quad
\text{Recall} = \frac{TP}{TP+FN}, \quad
\text{F1} = \frac{2 \cdot \text{Precision} \cdot \text{Recall}}{\text{Precision}+\text{Recall}},
\end{equation}
\begin{equation}
\text{AUC} = \int_{0}^{1} \text{TPR}(\text{FPR}) \, d\text{FPR},
\end{equation}
\begin{equation}
\text{FDR} = \frac{TP}{TP+FN}, \quad
\text{FAR} = \frac{FP}{FP+TN},
\end{equation}
\begin{equation}
\text{EDS} = \frac{1}{N}\sum_{i=1}^{N} \exp\left(-\gamma \cdot (t_i^{det}-t_i^{fraud})\right),
\end{equation}
\begin{equation}
\text{CDGS} = \frac{\text{F1}_{\text{target}}}{\text{F1}_{\text{source}}},
\end{equation}
where $TP$ denotes the number of correctly identified fraud samples, $FP$ denotes the number of normal samples incorrectly classified as fraud, $TN$ denotes the number of correctly identified normal samples, and $FN$ denotes the number of missed fraud samples. $\text{TPR}$ and $\text{FPR}$ represent the true positive rate and false positive rate, respectively. $N$ denotes the total number of test samples. $t_i^{fraud}$ and $t_i^{det}$ indicate the true occurrence time of the $i$-th fraud event and the first detection time predicted by the model, respectively. $\gamma$ is the temporal penalty coefficient. $\text{F1}_{\text{source}}$ and $\text{F1}_{\text{target}}$ represent the f1 scores achieved by the model on the source domain and target domain, respectively. These metrics reflect model performance from multiple perspectives. Specifically, auc and f1 measure overall discriminative capability, precision and recall focus on false alarm control and missed detection risk, fdr and far directly correspond to fraud identification accuracy and false alarm rates in practical business scenarios, early detection score emphasizes sensitivity to early weak fraud signals, and cross-domain generalization score evaluates the transferability and generalization capability across platforms, streamers, and product categories.

\subsection{{Overall Performance Comparison with Baseline Methods}%MDPI: This section heading and the captions of Figure 4 and Table 2 are repeat, please confirm whether it can be retained or should be revised.
%Authors: We have confirmed that it can be retained.
}

The objective of this experiment is to systematically evaluate, from an overall perspective, the comprehensive discriminative capability, risk recognition stability, and practical applicability of different categories of methods in the task of abnormal marketing detection in live-streaming e-commerce. By introducing representative baselines covering unsupervised methods, traditional temporal models, deep temporal models, graph-based models, and multimodal fusion models, the applicability boundaries of different modeling assumptions and mathematical structures in complex marketing scenarios are thoroughly examined.

As reported in Table~\ref{tab:baseline}, unsupervised methods based on distributional assumptions, including Isolation Forest and One-Class SVM, exhibit relatively inferior performance in terms of AUC and F1. This behavior can be attributed to the fact that such methods primarily rely on modeling a global normal distribution and identifying isolated deviations, whereas abnormal marketing behaviors in live-streaming scenarios often emerge gradually through weak signals and coordinated patterns, making them difficult to appear as isolated outliers in the feature space. As the modeling paradigm shifts from unsupervised distribution modeling to temporal modeling, LSTM and GRU achieve noticeable improvements in Recall, F1, and EDS, indicating that incorporating temporal dependency structures facilitates the capture of user behavior evolution. However, due to their reliance on limited-length hidden state propagation, recurrent models still suffer from gradient attenuation and constrained memory when handling long time spans and abrupt anomalies, which restricts their gains in AUC, EDS, and CDGS.
\begin{table}[H]
%\footnotesize
%\centering
\caption{{Overall} %MDPI: 1.Please confirm if the different arrows need to be explained, and please check all tables. 2.We moved Table 2 to AFTER where it is first mentioned in the text. Please confirm and check all tables.
%Authors: 1. We've confirmed, it's not needed. 2. We confirmed and checked.
 performance comparison with baseline methods.}
\label{tab:baseline}
\begin{adjustwidth}{-\extralength}{0cm}
%} % If the paper is ``preprints'', please uncomment this parenthesis.
%\isPreprints{\begin{tabularx}{\textwidth}{CCCC}}{% This command is only used for ``preprints''.
		\begin{tabularx}{\fulllength}{lcccCCCCC}
\toprule
\textbf{Method} & \textbf{AUC}~\boldmath{$\uparrow$} & \textbf{Precision}~\boldmath{$\uparrow$} & \textbf{Recall}~\boldmath{$\uparrow$} & \textbf{F1}~\boldmath{$\uparrow$} & \textbf{FDR}~\boldmath{$\uparrow$} & \textbf{FAR}~\boldmath{$\downarrow$} & \textbf{EDS}~\boldmath{$\uparrow$} & \textbf{CDGS}~\boldmath{$\uparrow$} \\
\midrule
Isolation Forest & 0.781 & 0.702 & 0.648 & 0.674 & 0.648 & 0.142 & 0.412 & 0.681 \\
One-Class SVM & 0.793 & 0.718 & 0.662 & 0.689 & 0.662 & 0.136 & 0.438 & 0.695 \\
LSTM & 0.834 & 0.756 & 0.721 & 0.738 & 0.721 & 0.121 & 0.502 & 0.734 \\
GRU & 0.842 & 0.764 & 0.728 & 0.746 & 0.728 & 0.118 & 0.517 & 0.742 \\
Transformer (temporal) & 0.867 & 0.791 & 0.756 & 0.773 & 0.756 & 0.106 & 0.561 & 0.771 \\
GNN-based Fraud Detection & 0.872 & 0.798 & 0.761 & 0.779 & 0.761 & 0.103 & 0.548 & 0.786 \\
MMBT & 0.883 & 0.812 & 0.774 & 0.793 & 0.774 & 0.097 & 0.584 & 0.801 \\
CLIP-style Fusion & 0.889 & 0.819 & 0.781 & 0.800 & 0.781 & 0.094 & 0.601 & 0.812 \\
\midrule
MM-FGDNet (Ours) & {\textbf{0.927} %MDPI: Please confirm if the bold formatting is necessary; if not, please remove it. The following highlights are the same.
%Authors: We have confirmed that it can be retained.
} & \textbf{{0.861}} & \textbf{{0.834}} & \textbf{{0.847}} & \textbf{{0.834}} & \textbf{{0.071}} & \textbf{{0.689}} & \textbf{{0.872}} \\
\bottomrule
\end{tabularx}
\end{adjustwidth}
\end{table}

\begin{figure}[H]
\hspace*{-.4em}\includegraphics[width=\linewidth]{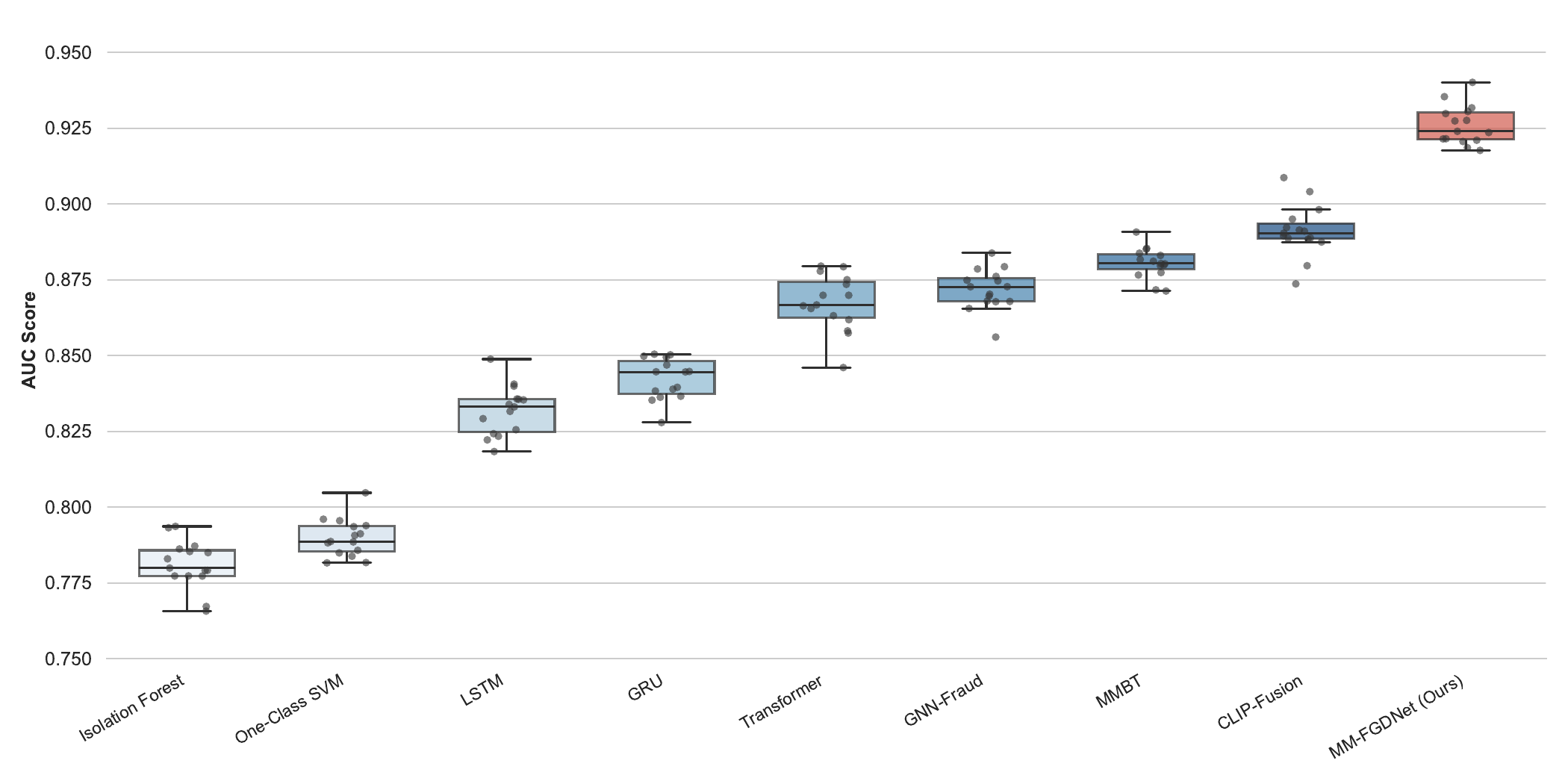}
\caption{{Overall} %MDPI: We moved Figure 4 to AFTER where it is first mentioned in the text. Please confirm..
%Authors: We confirmed.
 performance comparison with baseline methods.}
\label{fig: box}
\end{figure}

\vspace{-4pt}
As shown in Figure \ref{fig: box}, temporal Transformer models based on self-attention further enhance long-range dependency modeling and outperform recurrent models in AUC, F1, and EDS, highlighting the advantage of global temporal modeling for identifying abnormal trends. After introducing user relationship information at the structural level, GNN-based methods achieve further improvements in F1 and CDGS, demonstrating that group-structure modeling effectively captures user coordination in organized fraud behaviors. Nevertheless, these methods heavily depend on graph construction quality and make limited use of content modalities and temporal semantics, resulting in a performance bottleneck in early detection as reflected by EDS. Multimodal fusion models, including MMBT and CLIP-style fusion, further improve Precision, F1, and AUC by jointly modeling visual and textual features, confirming that cross-modal semantic complementarity alleviates the limitations of single-modality representations. However, such approaches often rely on static or weakly temporal fusion strategies and do not explicitly model temporal alignment and behavioral evolution across modalities, leaving room for improvement in FAR control and early risk perception. In contrast, MM-FGDNet consistently achieves superior performance across all evaluation metrics, with particularly pronounced advantages in F1, EDS, and CDGS. These gains stem from the systematic architectural design of the proposed framework. Cross-modal temporal alignment mitigates noise amplification caused by semantic misalignment, deep temporal modeling captures the progressive evolution from weak signals to abnormal bursts, and cooperative behavior modeling explicitly characterizes organized multi-user manipulation. By jointly constraining temporal, semantic, and structural dimensions, the proposed framework maintains a stable decision boundary under complex distributions and cross-scenario shifts, thereby significantly outperforming existing baselines in overall performance, early detection capability, and cross-domain generalization.
\vspace{-8pt}

\subsection{{Ablation Study of Different Components in MM-FGDNet}%MDPI: This section heading and the caption of Table 3 are repeat, please confirm whether it can be retained or should be revised.
%Authors: We have confirmed that it can be retained.
}

The ablation study is designed to systematically evaluate the individual contributions and collaborative effects of the core components in MM-FGDNet with respect to overall performance, early risk detection, and cross-scenario generalization. By removing the cross-modal temporal alignment module, the temporal fraud modeling module, the cooperative manipulation detection module, and the diffusion-based augmentation strategy one at a time while keeping the remaining architecture unchanged, the influence of different modeling assumptions on the decision boundary and risk perception capability can be directly analyzed.

As shown in Table~\ref{tab:ablation}, the full model consistently achieves the best performance across all metrics, indicating that the proposed components are not merely additive but form complementary constraints at both the mathematical structure and information flow levels. When the cross-modal alignment module is removed, both AUC and F1 decrease substantially, accompanied by a significant increase in FAR. This suggests that misalignment across modalities in temporal and semantic dimensions amplifies noise in the discriminative space, making non-fraudulent behaviors more likely to be misclassified as anomalies. This observation underscores the critical role of cross-modal alignment in reducing feature distribution shifts and stabilizing joint representations. When the temporal fraud modeling module is excluded, the EDS metric exhibits the most pronounced degradation, demonstrating a severe loss in sensitivity to early weak signals and validating the importance of explicitly modeling behavioral evolution for capturing the onset of fraudulent activities.

\begin{table}[H]
%\centering
\small
\caption{Ablation study of different components in MM-FGDNet.}
\label{tab:ablation}
\begin{tabularx}{\textwidth}{lCCCCCC}
\toprule
\textbf{Variant} & \textbf{AUC} $\uparrow$ &\textbf{F1}~\boldmath{$\uparrow$} & \textbf{FDR}~\boldmath{$\uparrow$} & \textbf{FAR}~\boldmath{$\downarrow$} & \textbf{EDS}~\boldmath{$\uparrow$} & \textbf{CDGS}~\boldmath{$\uparrow$} \\
\midrule
Full MM-FGDNet & \textbf{{0.927}} & \textbf{{0.847}} & \textbf{{0.834}} & \textbf{{0.071}} & \textbf{{0.689}} & \textbf{{0.872}} \\
\textit{w}/\textit{o} Cross-Modal Alignment & 0.898 & 0.812 & 0.798 & 0.089 & 0.602 & 0.821 \\
\textit{w}/\textit{o} Temporal Fraud Modeling & 0.883 & 0.799 & 0.782 & 0.093 & 0.531 & 0.814 \\
\textit{w}/\textit{o} Cooperative Detection & 0.892 & 0.807 & 0.791 & 0.086 & 0.578 & 0.793 \\
\textit{w}/\textit{o} Diffusion Augmentation & 0.901 & 0.818 & 0.804 & 0.082 & 0.614 & 0.839 \\
\bottomrule
\end{tabularx}
\end{table}

\begin{figure}[H]
\hspace*{-3.5em}\includegraphics[width=0.9\linewidth]{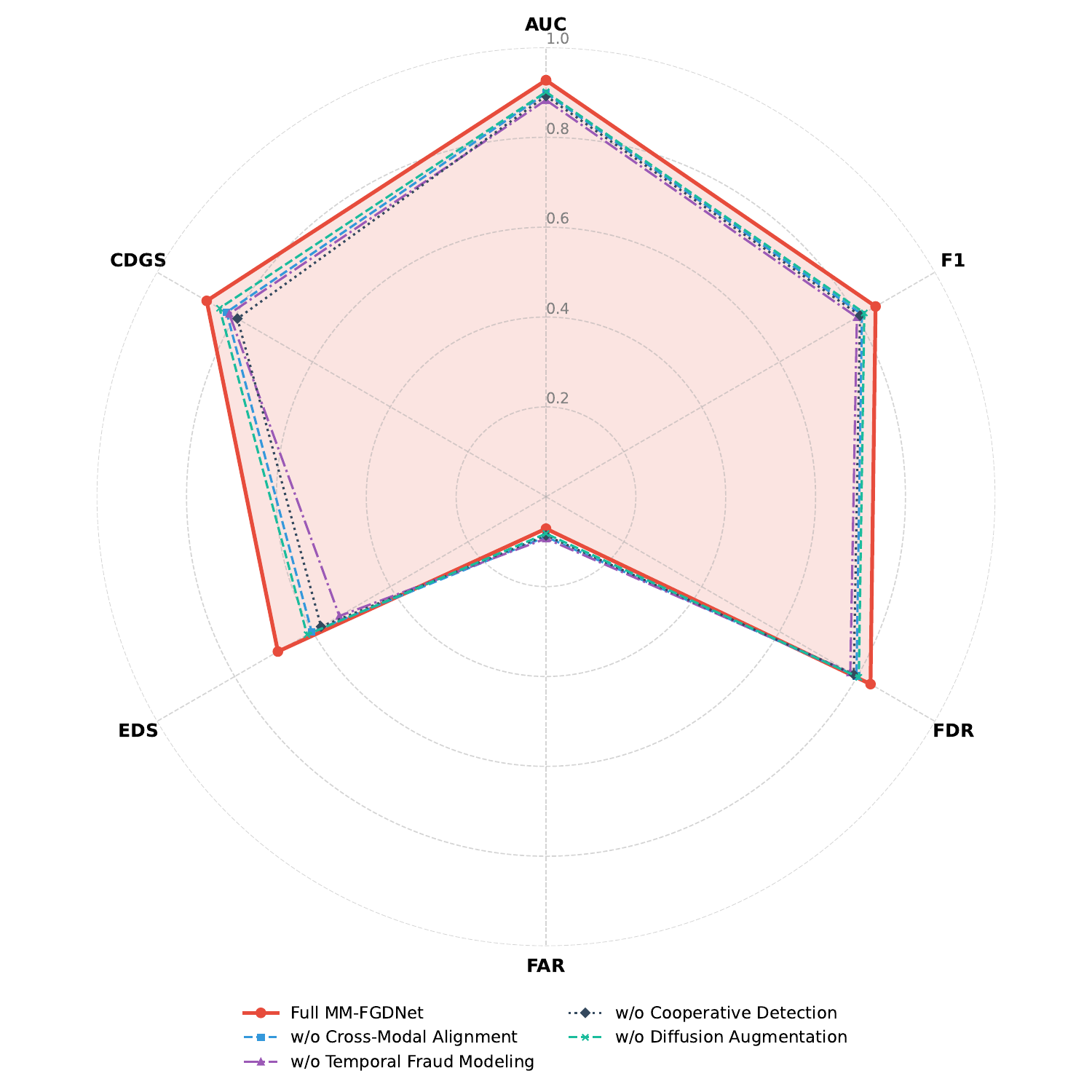}
\caption{{Radar chart illustrating the comprehensive ablation study results, comparing the full MM-FGDNet framework against four variants with specific core components removed. The axes represent six key evaluation metrics: AUC, F1-Score, Fraud Detection Rate (FDR), False Alarm Rate (FAR, plotted inversely where outward indicates lower FAR), Early Detection Score (EDS), and Cross-Domain Generalization Score (CDGS).}}
\label{fig: radar}
\end{figure}

{From the perspective of model structure, the performance variations observed across different ablation settings can be reasonably explained by their underlying mathematical modeling capacities. As shown in Figure \ref{fig: radar}, without temporal fraud modeling, the representation of the temporal dimension degenerates into near-static discrimination, causing the decision boundary to rely primarily on local feature distributions and limiting the accumulation of anomaly evidence across time steps. The removal of the cooperative detection module leads to a notable decline in CDGS indicating reduced generalization ability across streamers and platforms. This substantial performance gap can be attributed to the loss of high-order structural constraints among users; without these invariant topological patterns, the model is forced to rely more heavily on superficial platform- or individual-specific features rather than stable cooperative patterns. Although excluding diffusion-based augmentation results in relatively smaller performance drops, consistent declines in F1, EDS, and CDGS are still observed, suggesting that generative augmentation plays an important regularization role in smoothing the decision boundary and improving representation coverage under class imbalance and weak supervision. Overall, the ablation results validate the rationality of the MM-FGDNet design from both mathematical modeling capability and information constraint perspectives, demonstrating that cross-modal alignment reduces representation noise, temporal modeling accumulates anomaly evidence, and structural modeling constrains cooperative behaviors, collectively enabling stable, generalizable, and early-warning-capable fraud detection in complex marketing environments.}
\vspace{-8pt}

\subsection{{Cross-Domain Generalization Performance Under Different Target Scenarios}%MDPI: This section heading and the captions of Table 4 and Figure 6 are repeat, please confirm whether it can be retained or should be revised.
%Authors: We have confirmed that it can be retained.
}

The purpose of the cross-domain generalization experiment is to systematically evaluate the stability and transferability of different models under significant distribution shifts, thereby verifying whether abnormal marketing behaviors are captured at the mechanism level rather than relying on superficial characteristics such as streamer style, product attributes, or platform-specific rules. By constructing three target scenarios, namely a new streamer, a new product category, and a new platform, robustness is examined under conditions involving changes in content entities, variations in business semantics, and shifts in system environments.

As shown in Table~\ref{tab:cross}, the overall performance of all models declines under cross-domain conditions compared with in-domain settings, although the magnitude of degradation varies substantially. The temporal transformer-based model exhibits relatively stable yet limited generalization across all scenarios, with the most pronounced drops in $F1$ and $AUC$ observed in the new platform setting, indicating that reliance on temporal dependency modeling alone remains sensitive to platform-level distributional shifts. Multimodal fusion models achieve moderate improvements in the new streamer and new product scenarios by incorporating visual and textual information; however, the gains remain limited under the new platform condition, suggesting that static semantic fusion encounters adaptation bottlenecks when migrating across system environments. GNN-based methods consistently outperform the former two categories across all scenarios, particularly in terms of $F1$ and $AUC$, demonstrating that group-structure information exhibits stronger cross-domain stability. Nevertheless, the relatively limited improvement in $EDS$ indicates that such methods remain insufficiently sensitive to the early stages of abnormal behavior.

\begin{table}[H]
%\centering
\caption{Cross-domain generalization performance under different target scenarios.}
\label{tab:cross}
\begin{tabularx}{\textwidth}{lcCCC}
\toprule
\textbf{Target Scenario} & \textbf{Method} & \textbf{F1}~\boldmath{$\uparrow$} & \textbf{AUC}~\boldmath{$\uparrow$} & \textbf{EDS}~\boldmath{$\uparrow$} \\
\midrule
\multirow{4}{*}{New Streamer}
& Transformer (temporal) & 0.742 & 0.851 & 0.512 \\
& Multimodal Fusion (MMBT) & 0.769 & 0.873 & 0.548 \\
& GNN-based Fraud Detection & 0.779 & 0.881 & 0.536 \\
& MM-FGDNet & \textbf{{0.816}} & \textbf{{0.914}} & \textbf{{0.662}} \\
\midrule
\multirow{4}{*}{New Product Category}
& Transformer (temporal) & 0.736 & 0.846 & 0.498 \\
& Multimodal Fusion (MMBT) & 0.761 & 0.869 & 0.531 \\
& GNN-based Fraud Detection & 0.771 & 0.878 & 0.519 \\
& MM-FGDNet & \textbf{{0.808}} & \textbf{{0.909}} & \textbf{{0.643}} \\
\midrule
\multirow{4}{*}{New Platform}
& Transformer (temporal) & 0.721 & 0.832 & 0.471 \\
& Multimodal Fusion (MMBT) & 0.747 & 0.861 & 0.503 \\
& GNN-based Fraud Detection & 0.756 & 0.871 & 0.491 \\
& MM-FGDNet & \textbf{{0.801}} & \textbf{{0.902}} & \textbf{{0.628}} \\
\bottomrule
\end{tabularx}
\end{table}

\begin{figure}[H]
\includegraphics[width=\linewidth]{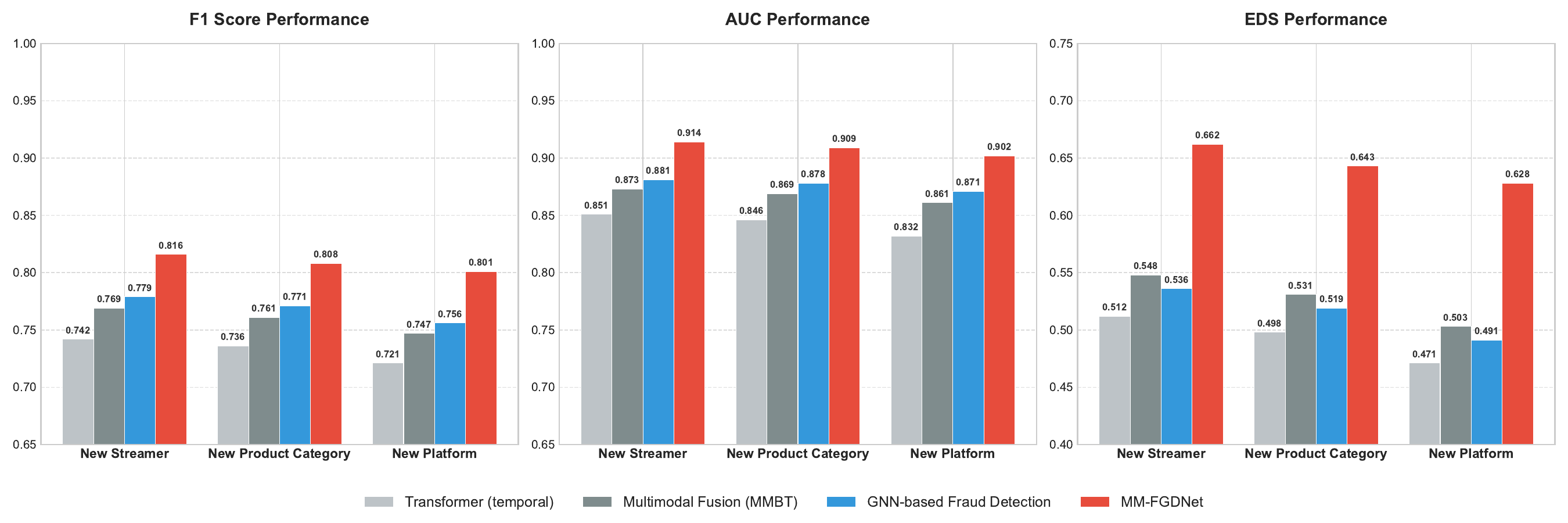}
\caption{Cross-domain generalization performance under different target scenarios.}
\label{fig: bar}
\end{figure}

As shown in Figure \ref{fig: bar}, the attention weights learned by temporal transformer models are optimized under source-domain distributions; when streamer pacing, product conversion dynamics, or platform interaction rules change, the learned temporal dependency patterns become misaligned, resulting in decision boundary drift. Although multimodal fusion models mitigate single-modality bias by introducing additional semantic views, the absence of explicit temporal alignment and cross-modal consistency constraints allows domain-specific noise to interfere with fused representations. GNN-based models benefit from user collaboration structures learned through graph propagation, which remain relatively invariant across scenarios, thereby enabling improved structural generalization. However, the de-emphasis on temporal evolution limits their ability to detect early weak signals. In contrast, MM-FGDNet consistently achieves superior performance across all three scenarios, with particularly notable advantages in $EDS$, indicating that abnormal behavior evolution can still be perceived at an early stage under distribution shifts. This robustness arises from the joint constraint mechanism embedded in the model architecture: cross-modal temporal alignment mitigates semantic drift across domains, temporal fraud modeling accumulates evolution evidence independent of specific domains, and cooperative manipulation detection introduces group-level structural priors as stable anchors. Consequently, the model maintains focus on the core generative mechanisms of abnormal marketing behaviors when facing changes in streamers, products, or platforms, thereby exhibiting stronger generalization capability and practical value.
\vspace{-4pt}
\subsection{Discussion}

The performance advantages demonstrated by MM-FGDNet across multiple experiments carry clear and concrete practical implications, particularly in real-world live-streaming e-commerce environments characterized by high dynamics and strong interactivity. In operational settings, abnormal marketing behaviors rarely manifest as abrupt and explicit violations; instead, risk often accumulates gradually through subtle changes in streamer behavior, comment sentiment, and user interaction rhythms. For instance, certain live-streaming sessions may appear compliant in terms of content and presentation, while the comment section suddenly exhibits highly homogeneous emotional expressions accompanied by unusually concentrated tipping activity. Such patterns are easily overlooked by manual inspection or rule-based systems. By performing cross-modal temporal alignment, MM-FGDNet maps content variations, emotional fluctuations, and transactional behaviors into a unified temporal semantic space, enabling systematic detection of these seemingly normal yet intrinsically abnormal early signals and providing platforms with more proactive risk warning capabilities. From the perspective of platform governance, the explicit modeling of cooperative behaviors offers interpretable evidence for identifying organized water-army groups and bot clusters. In practice, platforms frequently encounter scenarios in which individual user actions remain compliant while collective outcomes are anomalous, such as multiple newly registered accounts simultaneously tipping, liking, and commenting on the same product. While each individual action may appear legitimate, the overall interaction rhythm deviates significantly from that of normal user populations. Traditional models based on user profiling or point-wise anomaly detection struggle to address such cases. By modeling interaction structures and cooperative patterns among users, MM-FGDNet identifies consistency in timing, interaction targets, and emotional expression at the group level, revealing how abnormal popularity is orchestrated and amplified. This capability not only reduces the risk of mistakenly penalizing legitimate users but also provides actionable and interpretable insights to support fine-grained governance strategies.

Furthermore, the stable performance observed in cross-domain experiments highlights the strong deployment potential of the proposed framework across multiple platforms and product categories. In real business environments, platform rules, streamer profiles, and product structures evolve continuously, and models trained on historical data or single scenarios often require frequent retraining to maintain effectiveness. By modeling the generative mechanisms of abnormal marketing behaviors rather than memorizing platform-specific features, MM-FGDNet preserves detection capability when confronted with new streamers, new products, or even entirely new platforms. This property enables the framework to function as a core component of platform-level risk monitoring systems, supporting human review processes, optimizing recommendation strategies, and enhancing regulatory efficiency while safeguarding user experience and ecosystem fairness.

{Regarding system implementation and deployment, MM-FGDNet is designed to operate within a high-concurrency stream processing architecture suitable for large-scale platforms. To balance detection accuracy with real-time response requirements, a hierarchical deployment strategy is recommended. In this setup, lightweight rule-based filters serve as the first line of defense to screen out obvious traffic anomalies, while MM-FGDNet functions as the core deep analysis engine for processing complex, ambiguous interaction sequences. By leveraging distributed inference frameworks (e.g., TensorRT on GPU clusters) and asynchronous message queues, the model can process multimodal data streams with second-level latency, ensuring that risk warnings are generated before significant ecosystem damage occurs. Additionally, the interpretable subgraphs generated by the cooperative detection module can be directly integrated into human-in-the-loop review workstations, visualizing the core members of fraud syndicates to assist moderators in making rapid, evidence-based decisions.}

\subsection{Limitation and Future Work}

Although MM-FGDNet achieves substantial performance improvements in abnormal marketing detection for live-streaming e-commerce, several aspects remain open for further investigation. First, the joint modeling of multimodal signals and group-level structures enhances representational capacity but also incurs increased computational and storage overhead, which may pose challenges in ultra-large-scale platforms or latency-sensitive scenarios where careful trade-offs between accuracy and efficiency are required. Second, the cooperative manipulation detection module relies on the availability and quality of user interaction graphs; when interaction data are sparse or user relationships are partially unobservable due to platform constraints, detection effectiveness may be affected. Moreover, while cross-domain experiments demonstrate strong generalization, the model still primarily transfers knowledge from historical patterns when encountering extreme cold-start scenarios or entirely novel marketing mechanisms, which may lead to delayed responses. Future research may extend the framework along several directions. More lightweight cross-modal alignment and graph modeling strategies could be explored, potentially in combination with model compression or online learning techniques, to improve deployability in large-scale real-time environments. In addition, incorporating stronger causal modeling or active learning mechanisms may reduce reliance on historical fraud patterns and enhance adaptability to emerging abnormal behaviors. Finally, tighter integration between model outputs, platform governance rules, and human review workflows may facilitate the construction of human-in-the-loop risk identification and intervention systems, promoting sustained and effective real-world deployment.

\section{Conclusions}
{A data-driven multimodal abnormal behavior detection framework, termed MM-FGDNet, has been presented to address the growing concealment, coordination, and cross-modal heterogeneity of abnormal marketing behaviors in live-streaming e-commerce and digital marketing environments. By systematically modeling abnormal behavior mechanisms from the complementary perspectives of temporal evolution and group-level coordination, the proposed framework overcomes the limitations of rule-based or static-feature-driven approaches that are increasingly inadequate for large-scale platform governance and early risk warning. MM-FGDNet integrates cross-modal temporal alignment, temporal fraud pattern modeling, and cooperative manipulation detection into a unified architecture, enabling robust aggregation of weak signals across time and explicit characterization of organized group behaviors. Crucially, the framework incorporates interpretable abnormal subgraph analysis within the cooperative behavior detection module. By visualizing high-order interaction structures and pinpointing the core members of organized groups---such as bot clusters and paid poster networks---this mechanism provides granular, intuitive evidence that empowers human reviewers to rapidly validate algorithmic alerts, thereby significantly accelerating evidence-based decision-making in platform governance workflows. Extensive experiments on real-world multi-platform datasets demonstrate that MM-FGDNet consistently outperforms representative baseline methods, achieving an AUC of $0.927$ and an F1 score of $0.847$, with precision and recall reaching $0.861$ and $0.834$, respectively, while significantly reducing false alarm rates. Moreover, the proposed framework attains an Early Detection Score of $0.689$, indicating a superior capability to identify abnormal behaviors at their incipient stages. Ablation and cross-domain generalization studies further confirm the robustness and transferability of the proposed design across new streamers, new product categories, and new platforms. Overall, MM-FGDNet provides an effective, scalable, and practically deployable data-driven artificial intelligence solution for high-accuracy abnormal behavior detection and intelligent risk governance in live-streaming platforms.}

%%%%%%%%%%%%%%%%%%%%%%%%%%%%%%%%%%%%%%%%%%
\vspace{6pt} 

%%%%%%%%%%%%%%%%%%%%%%%%%%%%%%%%%%%%%%%%%%

\authorcontributions{Conceptualization, J.L. (Jingwen Luo), P.Z., Y.W. and Y.Z.; Data curation, J.L. (Jingqi Li) and X.K.; Formal analysis, Z.X.; Funding acquisition, Y.Z.; Investigation, Z.X.; Methodology, J.L. (Jingwen Luo), P.Z. and Y.W.; Project administration, Y.Z.; Resources, J.L. (Jingqi Li) and X.K.; Software, J.L. (Jingwen Luo), P.Z. and Y.W.; Supervision, Y.Z.; Validation, Z.X.; Visualization, J.L. (Jingqi Li) and X.K.; Writing---original draft, J.L. (Jingwen Luo), P.Z., Y.W., Z.X., J.L. (Jingqi Li), X.K. and Y.Z.; J.L. (Jingwen Luo), P.Z., and Y.W. contributed equally to this work. All authors have read and agreed to the published version of the manuscript}

\dataavailability{The data presented in this study are available on request from the corresponding author.}

\conflictsofinterest{The authors declare no conflict of interest.} 

%%%%%%%%%%%%%%%%%%%%%%%%%%%%%%%%%%%%%%%%%%
\begin{adjustwidth}{-\extralength}{0cm}

\reftitle{References}

% If authors have biography, please use the format below
%\section*{Short Biography of Authors}
%\bio
%{\raisebox{-0.35cm}{\includegraphics[width=3.5cm,height=5.3cm,clip,keepaspectratio]{Definitions/author1.pdf}}}
%{\textbf{Firstname Lastname} Biography of first author}
%
%\bio
%{\raisebox{-0.35cm}{\includegraphics[width=3.5cm,height=5.3cm,clip,keepaspectratio]{Definitions/author2.jpg}}}
%{\textbf{Firstname Lastname} Biography of second author}

% For the MDPI journals use author-date citation, please follow the formatting guidelines on http://www.mdpi.com/authors/references
% To cite two works by the same author: \citeauthor{ref-journal-1a} (\citeyear{ref-journal-1a}, \citeyear{ref-journal-1b}). This produces: Whittaker (1967, 1975)
% To cite two works by the same author with specific pages: \citeauthor{ref-journal-3a} (\citeyear{ref-journal-3a}, p. 328; \citeyear{ref-journal-3b}, p.475). This produces: Wong (1999, p. 328; 2000, p. 475)

%%%%%%%%%%%%%%%%%%%%%%%%%%%%%%%%%%%%%%%%%%
%% for journal Sci
%\reviewreports{\\
%Reviewer 1 comments and authors’ response\\
%Reviewer 2 comments and authors’ response\\
%Reviewer 3 comments and authors’ response
%}
%%%%%%%%%%%%%%%%%%%%%%%%%%%%%%%%%%%%%%%%%%
\PublishersNote{}
%\isPreprints{}{% This command is only used for ``preprints''.
\end{adjustwidth}
%} % If the paper is ``preprints'', please uncomment this parenthesis.
\end{document}